\documentclass[preprint]{ptephy_v1}%%%%%% to generate preprint number

\preprintnumber{XXXX-XXXX} %%% %%% Insert preprint number here

\usepackage{graphics} 
\usepackage{txfonts}
\usepackage{color}

\begin{document}

\title{Damping of the Higgs and Nambu-Goldstone modes of superfluid Bose gases at finite temperatures}

\author[1,*]{Kazuma Nagao}
\affil{Yukawa Institute for Theoretical Physics, Kyoto University, Kyoto 606-8502, Japan {\rm \email{kazuma.nagao@yukawa.kyoto-u.ac.jp}}}
\author[1]{Ippei Danshita} 

\begin{abstract}%
We study collective modes of superfluid Bose gases in optical lattices at commensurate fillings. We focus on the vicinity of the quantum phase transition to the Mott insulator, where there exists the Higgs amplitude mode in addition to the Nambu-Goldstone phase mode associated with the spontaneous U(1) symmetry breaking. We analyze finite-temperature effects on the damping of the collective modes by using an effective spin-1 model and the field theoretical methods based on the finite-temperature Green's function. We calculate the damping rates up to 1-loop order and evaluate them analytically and numerically. We show that the damping rate of the Higgs mode increases with increasing the temperature but it remains underdamped up to a typical temperature achieved in experiments. Moreover, we find that the Nambu-Goldstone mode attenuates via a Landau damping process resulting from interactions with the Higgs mode and it can be overdamped at the typical temperature in a certain parameter region.
\end{abstract}

\subjectindex{A63, I22}

\maketitle

\section{Introduction}%
In a system with spontaneous breaking of a continuous symmetry, there exists a gapless collective mode that corresponds to a motion restoring the symmetry, namely the Nambu-Goldstone mode~\cite{nambu-60, goldstone-61}. If the system also has a particle-hole symmetry, a gapful collective mode emerges. It corresponds to fluctuations of the order-parameter amplitude and is often referred to as the Higgs mode because of its analogy with the Higgs scalar boson in the particle physics~\cite{higgs-64}. In recent years, Higgs modes in condensed matter and ultracold gases have attracted particular attention~\cite{volovik-14, pekker-15} thanks to experimental developments for detecting those modes in various systems, including superconductors ${\rm NbSe_{2}}$ \cite{sooryakumar,littlewood,measson} and ${\rm Nb_{1-x}Ti_x N}$ \cite{matsunaga1,matsunaga2,sherman,comment}, quantum antiferromagnets $\rm TlCuCl_3$ \cite{ruegg,merchant-14} and $\rm KCuCl_3$ \cite{kuroe}, charge-density-wave materials ${\rm K_{0.3}MoO_3}$ \cite{demsar,schaefer} and ${\rm TbTe_3}$ \cite{yusupov,mertelj}, superfluid $\rm ^3He$ B-phase \cite{avenel,collett}, and superfluid Bose gases in optical lattices \cite{bissbort-11,endres-12}.

As for bosons in optical lattices, it has been predicted~\cite{altman-02,huber-07} that the Higgs mode emerges in the superfluid state near the quantum phase transition to the Mott insulator~\cite{greiner-02} at commensurate fillings, where the system has an approximate particle-hole symmetry~\cite{sachdev-11}. The most careful experimental analyses regarding the Higgs modes thus far have been made in Ref.~\cite{endres-12}; a response of two-dimensional (2D) gases to temporal modulation of the lattice amplitude has been measured as a function of the modulation frequency by means of the optical-lattice microscope techniques. Although the measured onset frequency of the response agrees with the theoretically computed energy gap of the Higgs mode, the response exhibits a broad continuum above the onset rather than a sharp peak. This means that the existence of the Higgs mode as a well-defined resonance has not yet been experimentally verified in this system.

The experiments of Ref.~\cite{endres-12} have triggered extensive theoretical investigations on the Higgs modes of 2D bosons in optical lattices~\cite{pollet-12,podolsky-12,gazit-13,chen-13,rancon-14,liu-15,katan-15}. In particular, previous studies using the quantum Monte Carlo simulations of the 2D Bose-Hubbard model~\cite{pollet-12, liu-15} have shown that the broadening of the spectral response is due to the combined effects of quantum fluctuations, finite temperatures, and spatial inhomogeneity caused by the confinement potential. Because effects of quantum and thermal fluctuations are in general weaker in higher dimensions, it is expected that 3D systems are advantageous over 2D for observing a resonance peak of the Higgs mode in experiments. Altman and Auerbach~\cite{altman-02} have indeed shown that the resonance peak of the Higgs mode in a homogeneous Bose-Hubbard system at zero temperature is significantly sharper in 3D than in 2D, i.e., the damping of the Higgs mode is weaker in 3D. As a next step, it is important to address effects of finite temperatures in 3D.

Another interesting respect of the Higgs mode is that its interaction with the NG mode may dramatically change properties of the latter mode. For instance, Nakayama {\it et al.}~\cite{nakayama} recently predicted the Fano resonance of the NG mode mediated by a bound Higgs mode localized around potential barriers as a result of coupling between the two modes induced by the barriers. At finite temperatures, dynamically excited NG modes can interact with thermally excited Higgs modes and one of the natural consequences of such interactions should be the damping of the NG mode.

In this paper, we study damping of the Higgs and NG modes of 3D superfluid Bose gases in homogeneous optical lattices in the vicinity of the Mott transition at commensurate fillings, with a particular focus on effects of finite temperatures. To calculate the damping rates, we use a low-energy effective model \cite{altman-02}, which has the same form as the $S=1$ XY model with the uniaxial single-ion isotropy, and the field theoretic approach based on the finite-temperature Green's function. We analytically obtain approximate expressions of the damping rates for the two modes and also present numerical evaluations of the damping rates to clarify the validity region of the analytical formulae. We show that the damping rate of the Higgs mode at zero momentum increases with increasing the temperature but it remains smaller than the oscillation frequency of the mode at finite temperatures that can be realized in typical experiments. Moreover, we find that the interactions between the two modes allow for the Landau damping process of the NG mode, in which the damping rate can be even larger than the mode frequency at the typical temperatures, i.e., the NG mode can be overdamped.

This paper is organized as follows. In Sec.~\ref{sec:model}, we briefly review the mapping of the Bose-Hubbard model onto the effective spin-1 model. In Sec.~\ref{sec:methods}, we explain our methods to calculate the damping rates at finite temperatures. In Sec.~\ref{sec:Higgs}, using the formula obtained in Sec.~\ref{sec:methods}, we calculate the damping rate of the Higgs mode to discuss its dependence on the temperature and the interaction strength. In Sec.~\ref{sec:NG}, we discuss properties of the damping of the NG mode. In Sec.~\ref{sec:summary}, the results are summarized. Through out the below discussion, we set $\hbar=k_{\rm B}=a=1$, where  $\hbar$, $k_{\rm B}$, and $a$ denote the reduced Plank constant, the Boltzmann constant, and the lattice constant.

%%%%%%%%%%%%%%%%%%
\section{Model}
\label{sec:model}
%%%%%%%%%%%%%%%%%%
We consider ultra-cold bosonic atoms in a cubic optical lattice. Assuming that the lattice is sufficiently deep, the system can be described by the Bose-Hubbard model within the tight-binding approximation \cite{fisher-89,jaksch-98},
\begin{equation}
H=-J\sum_{\langle i,j\rangle}(a_i^{\dagger}a_j + {\rm h.c.})+\frac{U}{2}\sum_{i}(a_i^{\dagger}a_i-{\bar n})^2 -\delta\mu\sum_i (a_i^{\dagger}a_i-{\bar n}), \label{bh1}
\end{equation}
where $a_i^{\dagger}$ and $a_i$ denote the creation and annihilation operators for a boson at site $i$ and satisfy the bosonic commutation relations $[a_i,a_j^{\dagger}]=\delta_{i,j}$, $[a_i,a_j]=[a_i^{\dagger},a_j^{\dagger}]=0$. Here, ${\bar n} \equiv \langle a_i^{\dagger}a_i \rangle$ is the filling factor. The symbol $\langle i,j\rangle$ denotes the nearest-neighbor sites. The parameters $J$, $U$, and $\delta \mu$ are the hopping energy, the on-site interaction energy, and the chemical potential. Note that at a sufficiently large filling ${\bar n} \gg 1$, $\delta \mu =0$ corresponds to the commensurate case, i.e. integer filling ${\bar n}\in{\bf N}$. This definition simplifies the following derivation of the effective model to describe the system near the Mott phase with filling $\bar n$. In this paper, we focus on the high filling case.

In the vicinity of the ${\bar n}$th Mott phase, only three states \{$|{\bar n}-1\rangle_i$, $|{\bar n}\rangle_i$, $|{\bar n}+1\rangle_i$\} per site dominate the low-energy behavior of the superfluid near the Mott phase because the local fluctuations from the mean-field ground state in the Mott phase $\prod_i|{\bar n}\rangle_i$ are sufficiently suppressed so that we can ignore the high-energy excited states $\{|{\bar n}\pm2\rangle_i, |{\bar n}\pm3\rangle_i, \cdots\}$ in the complete Hilbert space. In the reduced Hilbert subspace, the operators in Eq.~(\ref{bh1}) take an approximated form. In order to obtain the specific expressions, it is convenient to introduce Schwinger bosons \cite{schwinger1} in the following manner, 
\begin{align}
|{\bar n}+1\rangle_i = t_{1i}^{\dagger}|{\rm vac}\rangle,\hspace{2mm}|\bar n\rangle_i = t_{0i}^{\dagger}|{\rm vac}\rangle,\hspace{2mm}|{\bar n}-1\rangle_i = t_{-1i}^{\dagger}|{\rm vac}\rangle,
\label{sch}
\end{align}
where $|{\rm vac}\rangle$ is a vacuum of the Schwinger bosons. These operators satisfy the bosonic commutation relations  $[t_{mi},t_{nj}^{\dagger}]=\delta_{m,n}\delta_{i,j}$ $(m,n=-1,0,1)$ and we impose a local constraint $\sum_{m=-1}^{m=1} t_{mi}^{\dagger}t_{mi}=1$ on them to eliminate non-physical states, e.g. $t_{1i}^{\dagger}t_{0i}^{\dagger}|{\rm vac}\rangle$. The operator $t_{0i}$ represents the local mean field in the Mott phase and $t_{1i}$ ($t_{-1i}$) is the single particle (hole) excitaion at site $i$. The bosonic operator $a_i^{\dagger}$ can be expressed in terms with the Schwinger bosons (\ref{sch}),
\begin{align}
a_i^{\dagger}&=\sqrt{{\bar n}+1}t_{1i}^{\dagger}t_{0i}+\sqrt{\bar n}t_{0i}^{\dagger}t_{-1i} \label{a}.
\end{align}
Substituting these expressions into the Bose-Hubbard model (\ref{bh1}), we obtain the effective model which describes the low-energy behavior in the reduced space and has the same form as the XY model with the uniaxial single-ion isotropy and the magnetic coupling \cite{altman-02}, 
\begin{equation}
H_{\rm eff}=-\frac{J{\bar n}}{2}\sum_{\langle i,j \rangle}(S_{i}^{+}S_{j}^{-}+{\rm h.c.})+\frac{U}{2}\sum_{i}(S^{z}_i)^{2}-h\sum_iS_i^z,
\label{spin}
\end{equation}
where $h=\delta \mu$. The spin operators are written in terms with the Schwinger bosons (\ref{sch}),
\begin{equation}
S^{+}_{i}=\sqrt{2}(t_{1i}^{\dagger}t_{0i}+t_{0i}^{\dagger}t_{-1i}), \hspace{2mm} S^{-}_{i}=(S^{+}_{i})^{\dagger}, \hspace{2mm}  S^{z}_{i}=t_{1i}^{\dagger}t_{1i}-t_{-1i}^{\dagger}t_{-1i}.
\end{equation}
and obey the standard SU(2) algebra. For ${\bar n}Jz/U \gg 1$, the ground state is the XY ferromagnetic ordered state where the U(1) symmetry is spontaneously broken. Here $z=2d$ is the coordination number and $d=3$ is the spatial dimension of the system. For ${\bar n}Jz/U \ll 1$, the ground state has the U(1) symmetry and the mean-field wave function is given by $\prod_i | S=1,m_z=0\rangle_i \equiv \prod_i|{\bar n}\rangle_i$. Notice that in the case $h=\delta \mu =0$ the effective model (\ref{spin}) has the particle-hole symmetry, which corresponds to the one emerging in the model (\ref{bh1}) near the superfluid-Mott insulator transition at commensurate fillings. In fact, except the last term, the effective model (\ref{spin}) is invariant under the interchange of particle excitations $t_{1i}^{\dagger}$ ($t_{1i}$) and hole excitations $t_{-1i}^{\dagger}$ ($t_{-1i}$).

%%%%%%%%%%%
\section{Methods}
\label{sec:methods}
%%%%%%%%%%%
%%
\subsection{Canonical transformation}
Let us explain how to describe the collective excitations of the strongly correlated superfluid by using the effective model (\ref{spin}), within a mean-field approximation developed by Altman and Auerbach \cite{altman-02}. In the following discussion, we deal with the particle-hole symmetric case, i.e. $h=\delta \mu =0$. First, we define a canonical transformation as
\begin{align}
b_{0i}&={\rm cos}(\theta_0/2)t_{0i}+{\rm sin}(\theta_0/2)(t_{1i}+t_{-1i})/\sqrt{2}, \nonumber \\
b_{1 i}&= {\rm sin}(\theta_0/2)t_{0i}-{\rm cos}(\theta_0/2)(t_{1i}+t_{-1i})/\sqrt{2}, \label{b} \\
b_{2 i}&= (t_{1i}-t_{-1i})/\sqrt{2}, \nonumber
\end{align}
which generates a rotation of the old basis spanned by three Schwinger bosons ($t_{1i},t_{0i},t_{-1i}$) into a new basis. The new operators also satisfy the bosonic commutation relations $[b_{mi},b^{\dagger}_{nj}]=\delta_{m,n}\delta_{i,j}$, $[b_{mi},b_{nj}]=[b^{\dagger}_{mi},b^{\dagger}_{nj}]=0$ and the local constraint $\sum_{m=0}^{m=2}b^{\dagger}_{mi}b_{mi}=1$. The angle of transformation is given by $\theta_0={\rm tan}^{-1}(\sqrt{1-u^2}/u)$, where $u\equiv U/(4J\bar{n}z)$ is a dimensionless parameter. The transformation (\ref{b}) can be derived by the bosonic Gutzwiller mean-field ansatz in the reduced space~\cite{altman-02,huber-07}. The operators $b_{0i}$ and  $b_{1i}$ ($b_{2i}$) stand for the mean-field ground state and the excitations in the Higgs (NG) branch. Note that in the mean-field theory and the high-filling limit ${\bar n} \gg 1$, the superfluid to Mott insulator transition occurs at $u_c=1$. In other words, a quantity $|u-u_c|=|u-1|$ measures the distance from the transition point.

\subsection{Holstein--Primakoff expansion}
We assume that all of the fluctuations from the mean-field ground state are small. Then, the effective model (\ref{spin}) represented by the new basis $b_{mi}$ $(b_{mi}^{\dagger})$ $(m=0,1,2)$ can be simplified by employing a Holstein--Primakoff expansion \cite{holsteinprimakoff}. With the local constraint $\sum_{m=0}^{m=2}b^{\dagger}_{mi}b_{mi}=1$, this expansion eliminates $b_{0i}$ $(b^{\dagger}_{0i})$ from the model (\ref{spin}) as
\begin{align}
b_{m}^{\dagger}b_{0}&=b_{m}^{\dagger}\sqrt{1-b_{1}^{\dagger}b_{1}-b_{2}^{\dagger}b_{2}},\nonumber \\ 
&\approx b_{m}^{\dagger}-\frac{1}{2}b_{m}^{\dagger}b_{1}^{\dagger}b_{1}-\frac{1}{2}b_{m}^{\dagger}b_{2}^{\dagger}b_{2} +\cdots, \label{hp}
\end{align}
where $m=\{1,2\}$. The higher order terms are ignored if we take into account the leading vertex terms with third order with respect to the excitations. This  approximation is similar to the spin-wave expansion in the localized spin systems with long-range orders (more details can be found, e.g., in Ref. \cite{auerbach}). Notice that in the case of finite temperatures the above truncation is valid only at $d\ge3$. At lower dimensions, the thermal fluctuations are so strong that they destroy the long-range order of the mean-field ground state~\cite{merminwagner,hohenberg,coleman} and one can not justify ignorance of the higher terms in the expansion (\ref{hp}).

Substituting the expansion (\ref{hp}) into the model (\ref{spin}), we obtain the simplified model, which has the sequent terms with each order, and would show them up to the third order,
\begin{equation}
H_{\rm eff}\equiv H_{\rm eff}^{(0)}+H_{\rm eff}^{(1)}+H_{\rm eff}^{(2)}+H_{\rm eff}^{(3)}+\cdots, \label{eq:eff2}
\end{equation}
where the index $l$ in $H_{\rm eff}^{(l)}$ means $l$th order with respect to the operators $b_{mi}^{\dagger}$ and $b_{mi}$ ($m=1,2$). Here, we perform the Fourier transformation of the operators defined as
\begin{equation}
b_{mi}^{\dagger}=\frac{1}{\sqrt N}\sum_{{\bf k}}b_{m{\bf k}}^{\dagger}e^{-i{\bf x}_i \cdot {\bf k}}, \hspace{3mm}b_{mi}=\frac{1}{\sqrt N}\sum_{{\bf k}}b_{m{\bf k}}e^{i{\bf x}_i \cdot {\bf k}},\label{fourier}
\end{equation}
where $N$ is the system size and the vector ${\bf x}_i$ denotes site $i$. The summation with respect to momentum $\bf k$ is taken over the first Brillouin zone. Then, each term $H_{\rm eff}^{(l)}$ can be written as the following equations, respectively,
\begin{equation}
H_{\rm eff}^{(0)}=N\left(\frac{U}{2}s^2-2J{\bar n}zs^2c^2\right),\hspace{5mm}H_{\rm eff}^{(1)}= 2J{\bar n}zsc\sqrt{N}\left(c^2-s^2-u \right)(b_{1 {\bf 0}}^{\dagger}+b_{1 {\bf 0}})=0,
\end{equation}
\begin{align}
H_{\rm eff}^{(2)}&=\sum_{\bf k}B_{1}({\bf k})b_{1 {\bf k}}^{\dagger}b_{1 {\bf k}}+\sum_{\bf k}B_{2}({\bf k})(b_{1 {\bf k}}^{\dagger}b_{1 - {\bf k}}^{\dagger}+b_{1 {\bf k}}b_{1 -{\bf k}}+b_{1 {\bf k}}b_{1 {\bf k}}^{\dagger}+b_{1 {\bf k}}^{\dagger}b_{1 {\bf k}}) \nonumber \\
&+\sum_{\bf k}C_{1}({\bf k})b_{2 {\bf k}}^{\dagger}b_{2 {\bf k}}+\sum_{\bf k}C_{2{}}({\bf k})(b_{2 {\bf k}}^{\dagger}b_{2 - {\bf k}}^{\dagger}+b_{2 {\bf k}}b_{2 -{\bf k}}-b_{2 {\bf k}}b_{2 {\bf k}}^{\dagger}-b_{2 {\bf k}}^{\dagger}b_{2 {\bf k}}),
\label{eq:h2}
\end{align}
\begin{align}
H_{\rm eff}^{(3)}&=\sum_{\bf k_1}\sum_{\bf k_2}\sum_{\bf k_3}D_{1}({\bf k}_1,{\bf k}_2,{\bf k}_3)(b_{2 {\bf k_1}}^{\dagger}b_{2 {\bf k_2}}b_{1 {\bf k_3}}+{\rm h.c.}) \nonumber  \\
&+\sum_{\bf k_1}\sum_{\bf k_2}\sum_{\bf k_3}D_{2}({\bf k}_1,{\bf k}_2,{\bf k}_3)(b_{1 {\bf k_1}}^{\dagger}b_{1 {\bf k_2}}b_{1 {\bf k_3}}+{\rm h.c.}) \nonumber \\
&+\sum_{\bf k_1}\sum_{\bf k_2}\sum_{\bf k_3}D_{3}({\bf k}_1,{\bf k}_2,{\bf k}_3)(b_{1 {\bf k_1}}^{\dagger}b_{2 {\bf k_2}}b_{2 {\bf k_3}}-b_{2 {\bf k_1}}b_{1 {\bf k_2}}^{\dagger}b_{2 {\bf k_3}}^{\dagger}+{\rm h.c.}), 
\label{eq:h3}
\end{align}
where $c\equiv{\rm cos}(\theta_0/2)$ and $s\equiv{\rm sin}(\theta_0/2)$. The explicit expressions of some coefficients in Eqs.~(\ref{eq:h2}) and (\ref{eq:h3}) are given by
\begin{align}
B_{1}({\bf k})=2J{\bar n}z,&\hspace{3mm}B_{2}({\bf k})=-\frac{J{\bar n}z}{2}u^2\gamma_{\bf k}\nonumber \\
C_{1}({\bf k})=J{\bar n}z\left(1+u \right),&\hspace{3mm}C_{2}({\bf k})=\frac{J{\bar n}z}{4}(1+u) \gamma_{\bf k}\nonumber \\
D_{1}({\bf k}_1,{\bf k}_2,{\bf k}_3)=&-\frac{1}{\sqrt{N}}J{\bar n}zu\sqrt{1-u^2}\gamma_{{\bf k}_3}\delta_{{\bf k}_1-{\bf k}_2+{\bf k}_3}\nonumber \\
D_{2}({\bf k}_1,{\bf k}_2,{\bf k}_3)=&-\frac{1}{\sqrt{N}}2J{\bar n}zu\sqrt{1-u^2}\gamma_{{\bf k}_3}\delta_{{\bf k}_1-{\bf k}_2+{\bf k}_3} \nonumber\\
D_{3}({\bf k}_1,{\bf k}_2,{\bf k}_3)=&\frac{1}{\sqrt{N}}\frac{J{\bar n}z}{2}\sqrt{1-u^2}\gamma_{{\bf k}_3}\delta_{{\bf k}_1-{\bf k}_2-{\bf k}_3} \nonumber
\end{align}
where $\gamma_{\bf k}=\sum_{\bf e}e^{-i{\bf k}\cdot{\bf e}}/z=\sum_{s=1}^d{\rm cos}(k_s)/d$.

\subsection{Bogoliubov transformation}%
To diagonalize $H_{\rm eff}^{(2)}$, let us perform
a Bogoliubov transformation \cite{pitaevskii} in each branch defined as 
\begin{align}
b_{m{\bf k}}&=u_{m{\bf k}}\beta_{m{\bf k}}+v_{m-{\bf k}}^*\beta_{m-{\bf k}}^{\dagger},  \nonumber \\
b_{m-{\bf k}}^{\dagger}&=u_{m-{\bf k}}^*\beta_{m-{\bf k}}^{\dagger}+v_{m{\bf k}}\beta_{m{\bf k}}, \label{bogo}
\end{align}
where $m=\{1,2\}$. The new operators obey the same commutation relations $[\beta_{m{\bf p}},\beta^{\dagger}_{n{\bf q}}]=\delta_{n,m}\delta_{{\bf p},{\bf q}}$ as the previous operator such that the coefficients $u_{m{\bf k}}$ and $v_{m {\bf k}}$ satisfy a relation $|u_{m{\bf k}}|^2-|v_{m{\bf k}}|^2=1$. Assuming that $u_{m{\bf k}}$ and $v_{m {\bf k}}$ are real numbers and imposing the condition that $H_{\rm eff}^{(2)}$ in the new representation has no anomalous term, we determine the coefficients,
\begin{align}
u_{1 {\bf k}} &= \sqrt{\frac{2-u^2\gamma_{\bf k}}{4\sqrt{1-u^2\gamma_{\bf k}}}+\frac{1}{2}},\hspace{1mm}v_{1 {\bf k}} = \sqrt{\frac{2-u^2\gamma_{\bf k}}{4\sqrt{1-u^2\gamma_{\bf k}}}-\frac{1}{2}}, \label{uv1} \\
u_{2 {\bf k}} &= \sqrt{\frac{2-\gamma_{\bf k}}{4\sqrt{1-\gamma_{\bf k}}}+\frac{1}{2}},\hspace{1mm}v_{2 {\bf k}} = -{\rm sgn}(\gamma_{\bf k})\sqrt{\frac{2-\gamma_{\bf k}}{4\sqrt{1-\gamma_{\bf k}}}-\frac{1}{2}},\label{uv2}
\end{align}
where ${\rm sgn}(x)$ is the sign function. After the Bogoliubov transformation, Eq.~(\ref{spin}) reads
\begin{align}
H_{\rm eff}=\sum_{m=1}^{2}\sum_{{\bf k}}{\cal E}_{m\bf k}\beta_{m{\bf k}}^{\dagger}\beta_{m{\bf k}}+H_{\rm eff}^{(3)}. \label{dia}
\end{align}
The operator $\beta^{\dagger}_{m{\bf k}}$ creates a quasi-particle with the excitation energy ${\cal E}_{m\bf k}$, which is given by
\begin{align}
{\cal E}_{1\bf k}&=2J{\bar n}z\sqrt{1-u^2\gamma_{\bf k}},\label{hg}\\
{\cal E}_{2\bf k}&=J{\bar n}z(1+u)\sqrt{1-\gamma_{\bf k}}\label{ng}. 
\end{align}
It is easily seen that the dispersion of the Higgs mode  ${\cal E}_{1\bf k}$ has a energy gap $\Delta \equiv 2J{\bar n}z\sqrt{1-u^2}$ at zero momentum while that of the NG mode ${\cal E}_{2\bf k}$ is gapless. Obviously, the gap of the Higgs mode closes at the critical point $u=1$.

\begin{figure}[htbp]
\begin{center}
\includegraphics[width=140mm]{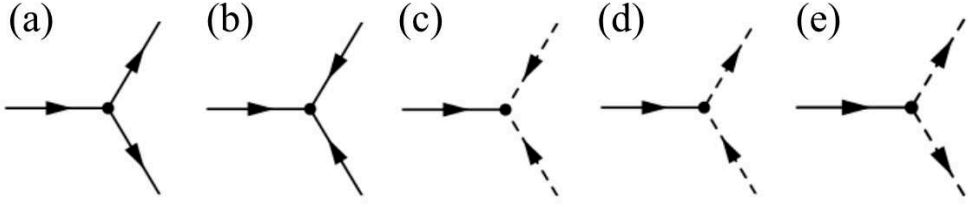}
\end{center}
\vspace{-5mm}
\caption{Independent interaction terms contained in the $H_{\rm eff}^{(3)}$. The solid and dashed lines represent the propagator of the Higgs mode and that of the NG mode. The incoming lines into and outgoing lines from the vertex correspond to the annihilation and creation of the quasiparticles.}
\label{vertex}
\end{figure}

Equation (\ref{dia}) has the third order terms characterizing the interactions among the three excitations. Specifically, $H_{\rm eff}^{(3)}$ has five types of interaction term, which are shown in Fig.~\ref{vertex}, and their Hermite conjugates. The interaction terms shown in Figs.~\ref{vertex}(a) and (b) generate scattering processes closed only in the Higgs branch while the terms shown in Figs.~\ref{vertex}(c), (d), and (e) couple the two different branches with one Higgs mode and two NG modes. As we will see in the following subsection, these terms cause damping of the elementary excitations. It is worth noting that $H_{\rm eff}^{(3)}$ does not include vertices consisting only of the propagators of the NG mode. This is a generic property of a superfluid with particle-hole symmetry~\cite{podolsky-11} and is in stark contrast to the case of a weakly interacting Bose gas, which has such vertices~\cite{beliaev, tsuchiya}.

%-----------------------------------------------------------------
%-----------------------------------------------------------------

When $|{\bf k}| \ll 1$, the above expressions are approximated with simpler forms. The simplification is done by the Taylor expansion of $\gamma_{\bf k}$ with respect to $\bf k$. For example, the excitation energies (\ref{hg}) and (\ref{ng}) become
\begin{align}
{\cal E}_{1\bf k}&\approx\sqrt{\Delta^2+c_{\rm h}^2|{\bf k}|^2}, \label{ahg} \\
{\cal E}_{2\bf k}&\approx c_{\rm ng}|{\bf k}|, \label{ang}
\end{align}
where $c_{\rm h}=2J{\bar n}\sqrt{z}u$ and $c_{\rm ng}=J{\bar n}\sqrt{z}(1+u)$. Similarly, if we assume that $u \neq u_c$, the coefficients (\ref{uv1}) and (\ref{uv2}) become
\begin{align}
u_{1 {\bf k}} \approx \sqrt{\frac{2-u^2+\Delta}{2\Delta}} + {\cal O}(k^2),&\hspace{5mm} v_{1 {\bf k}} \approx \sqrt{\frac{2-u^2-\Delta}{2\Delta}} + {\cal O}(k^2), \label{auv1}\\
u_{2 {\bf k}} \approx \sqrt{\frac{u+1}{4c_{\rm ng}k}}\left( 1+\frac{c_{\rm ng}k}{u+1} + {\cal O}(k^2) \right),&\hspace{5mm} v_{2 {\bf k}} \approx -\sqrt{\frac{u+1}{4c_{\rm ng}k}}\left( 1-\frac{c_{\rm ng}k}{u+1} + {\cal O}(k^2) \right).\label{auv2}
\end{align}
These expressions are isotropic in the momentum space, thus allowing for analytical evaluations of the damping rates.

\subsection{Finite-temperature Green's function}%%
To calculate the damping rates of the Higgs and the NG modes, we use the field-theoretical approaches based on the finite-temperature Green's function (See Ref. \cite{abrikosov,popov,negeleorland}). In our calculation, we define the Green's functions of the Higgs ($m=1$) and the NG ($m=2$) modes as
\begin{align}
\langle \beta_{m{\bf k}}(i \omega_n){\bar \beta_{m{\bf k}}}(i \omega_n) \rangle \equiv \frac{\int {\cal D}(\beta,{\bar \beta})\beta_{m{\bf k}}(i \omega_n){\bar \beta_{m{\bf k}}}(i \omega_n){\rm exp}(-{\cal S}_{\rm eff})}{\int  {\cal D}(\beta,{\bar \beta}){\rm exp}(-{\cal S}_{\rm eff})} ,
\end{align}
where $\omega_n=2\pi n/\beta$ $(n=0,\pm1,\pm2,\cdots)$ is the Matsubara frequency, $\beta = T^{-1}$ is the inverse temperature, $\beta_{m{\bf k}}(i \omega_n)$ and the conjugate ${\bar \beta_{m{\bf k}}}(i \omega_n)=(\beta_{m{\bf k}}(i \omega_n))^*$ are complex-valued field variables at $(\omega_n,{\bf k})$, and ${\cal D}(\beta,{\bar \beta})$ is a measure of the integrations. The effective action ${\cal S}_{\rm eff}={\cal S}^{(2)}_{\rm eff}+{\cal S}^{(3)}_{\rm eff}$ is derived from the effective model (\ref{dia}) and the quadratic action ${\cal S}^{(2)}_{\rm eff}$ is given by 
\begin{align}
{\cal S}_{{\rm eff}}^{(2)} &= \sum_{m=1}^{2}\sum_{n}\sum_{\bf k} (-i\omega_n+{\cal E}_{m\bf k}){\bar \beta_{m{\bf k}}}(i \omega_n)\beta_{m{\bf k}}(i \omega_n).
\end{align}
The third-order acton ${\cal S}^{(3)}_{\rm eff}$ can be obtained from $H_{\rm eff}^{(3)}$ with mere replacement of the operators with the field variables. 

In the field theory, the damping rates can be calculated as imaginary parts of the self energies of the Higgs mode $\Sigma_1(i\omega_n;{\bf k})$ and of the NG mode $\Sigma_2(i\omega_n;{\bf k})$. We define them as $\Gamma_{m{\bf k}}\equiv {\rm Im}\Sigma_{m}({\cal E}_{m\bf k}+i\epsilon;{\bf k})$ where $\epsilon$ is a positive-valued infinitesimal quantity. Calculating these damping rates by perturbative expansion with respect to ${\cal S}_{{\rm eff}}^{(3)}$ up to the second order, we obtain 
\begin{align}
\Gamma_{{1}{\bf k}} &=\frac{\pi}{2} \sum_{{\bf k}_1} \sum_{{\bf k}_2} | {\cal M}_{{\bf k},{\bf k}_1,{\bf k}_2} |^2 (1+f_{\rm B}({\cal E}_{2{\bf k}_1})+f_{\rm B}({\cal E}_{2{\bf k}_2}))\delta({\cal E}_{1\bf k}-{\cal E}_{2{\bf k}_1}-{\cal E}_{2{\bf k}_2}), \label{damp} \\
\Gamma_{2{\bf k}}&=\pi \sum_{{\bf k}_1} \sum_{{\bf k}_2}|{\cal M}_{{\bf k}_1,{\bf k},{\bf k}_2}|^2(f_B({\cal E}_{2\bf k_2})-f_B({\cal E}_{1\bf k_1})) \delta({\cal E}_{2\bf k}-{\cal E}_{1\bf k_1}+{\cal E}_{2\bf k_2}), \label{damp2}
\end{align}
where $\delta(x)$ is the $\delta$ function, $f_{\rm B}(x)=1/(e^{\beta x}-1)$ is the Bose distribution function, and the matrix elements ${\cal M}_{{\bf k}_1,{\bf k}_2,{\bf k}_3}$ are given by
\begin{align}
{\cal M}_{{\bf k}_1,{\bf k}_2,{\bf k}_3}=\frac{1}{\sqrt {N}} \delta_{{\bf k}_1,{\bf k}_2+{\bf k}_3}&\left[-J{\bar n}zu\sqrt{1-u^2}\gamma_{{\bf k}_1}(u_{1 {\bf k}_1}+v_{1 {\bf k}_1})(u_{2 {\bf k}_2}v_{2 {\bf k}_3} + v_{2 {\bf k}_2}u_{2 {\bf k}_3})\right. \nonumber  \\
&+\frac{J{\bar n}z}{2}\sqrt{1-u^2}\gamma_{{\bf k}_2}(u_{1 {\bf k}_1}u_{2 {\bf k}_3}-v_{1 {\bf k}_1}v_{2 {\bf k}_3} )(u_{2 {\bf k}_2} - v_{2 {\bf k}_2} ) \label{mat} \\
&+\left. \frac{J{\bar n}z}{2}\sqrt{1-u^2}\gamma_{{\bf k}_3}(u_{1 {\bf k}_1}u_{2 {\bf k}_2}-v_{1 {\bf k}_1}v_{2 {\bf k}_2})(u_{2 {\bf k}_3} - v_{2 {\bf k}_3} ) \right]. \nonumber
\end{align}

\begin{figure}[htbp]
\begin{center}
\includegraphics[width=100mm]{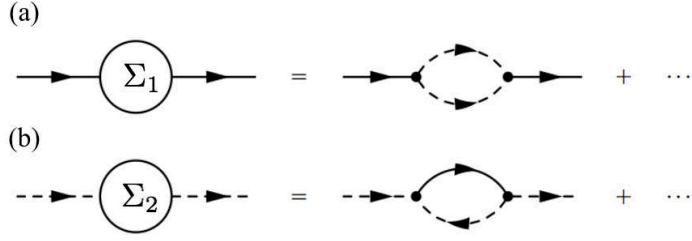}
\end{center}
\vspace{-5mm}
\caption{Contribution to the self energy of (a) the Higgs mode $\Sigma_{1}(i\omega_n;{\bf k})$ and (b) the NG mode $\Sigma_{2}(i\omega_n;{\bf k})$. Other diagrams don't yield the nonzero contribution of the damping rates.}
\label{fey}
\end{figure}

The contributions to each damping rate consist of only one type of the perturbative correction.
Figure {\ref{fey}} shows the Feynman diagrams that provide the non-zero contributions to the damping rates (\ref{damp}) and (\ref{damp2}). In general, there are other diagrams with different structures from the ones shown in Fig.~\ref{fey}. However, we find that the actual contributions to the damping come from only the diagrams depicted in Fig.~\ref{fey} within the 1-loop order, because of the energy-momentum conservations laws, which is represented as $\delta$ function in the equations (\ref{damp}) and (\ref{damp2}).

%-----------------------------------------------------------------
%-----------------------------------------------------------------

The diagram shown in Fig. {\ref{fey}}(a) means that the Higgs mode attenuates by decaying into the two NG modes due to the quantum and thermal fluctuations. This type of damping of collective modes is called  Beliaev damping \cite{beliaev,landau9}. In general, the Beliaev damping can occur even at zero temperature.
The Beliaev damping of the Higgs mode at zero temperature is first predicted by Altman and Auerbach \cite{altman-02}. On the other hand, the diagram depicted in Fig. {\ref{fey}}(b) means that the NG mode attenuates through processes in which the initial NG mode absorbs another NG mode excited thermally, and subsequently turns into the Higgs mode as the final state. This type of damping is called Landau damping \cite{landau10,pethick}. At zero temperature, this damping cannot occur because there is no thermal excitation. Therefore, in the vicinity of the Mott phase, the NG mode does not attenuate at zero temperature within the approximations discussed above. 

%%%%%%%%%%%%%%%%%%%%%%%%%%%%%%%
\section{Damping rate of the Higgs mode}
\label{sec:Higgs}
%%%%%%%%%%%%%%%%%%%%%%%%%%%%%%%
In this section, we discuss the damping of the Higgs mode at finite temperatures by evaluating the expression (\ref{damp}). In particular, we consider the case of the Higgs mode with zero momentum because a typical perturbation used for exciting the Higgs mode in experiments is the lattice-amplitude modulation with zero momentum~\cite{endres-12}.

First, let us evaluate the integrations of the formula (\ref{damp}) within a long-wavelength approximation, where ${\cal E}_{1{\bf k}}$, ${\cal E}_{2{\bf k}}$, $u_{m{\bf k}}$, and $v_{m{\bf k}}$ are approximated with Eqs.~(\ref{ahg}), (\ref{ang}), (\ref{auv1}), and (\ref{auv2}). This approximation is better justified in a closer vicinity of the critical point, $u=u_c$, where the energy of the NG mode dominant to the damping of the Higgs mode, $\Delta/2$, is smaller. Integrating with respect to ${\bf k}_1$ and ${\bf k}_2$ in the r.h.s. of Eq.~(\ref{damp}) within the approximation, we obtain a simple formula
\begin{align}
\Gamma_{1{\bf k}={\bf 0}}=\frac{3^{3/2}J{\bar n}z}{2^3\sqrt{2}\pi}(1+u)\sqrt{1-u^2}{\rm coth}\frac{\beta \Delta}{4}. \label{result1}
\end{align}
It is obvious from Eq.~(\ref{result1}) that the dependence on temperature $T$ is determined by the factor ${\rm coth}(\beta\Delta/4)$. At $T=0$ and $|u-u_c|\ll 1$, this factor becomes $1$ and the damping rate behaves as $\Gamma_{1{\bf k}={\bf 0}}\sim J{\bar n}z\sqrt{1-u^2}$. This consequence agrees with the previous one obtained by Altman and Auerbach \cite{altman-02}. In this sense, the analytic expression (\ref{result1}) gives the finite-temperature correction to the previous result at zero temperature. 

\begin{figure}[htbp]
\begin{center}
\includegraphics[width=80mm]{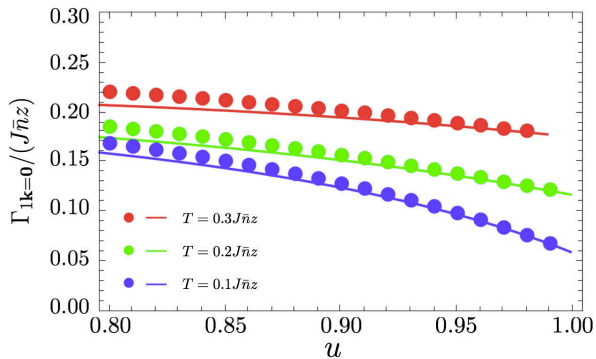}
\end{center}
\vspace{-5mm}
\caption{Dependence on $u$ of the damping rate $\Gamma_{1{\bf k}={\bf 0}}/(J\bar{n}z)$ at fixed temperatures. The solid lines represent the analytic expression (\ref{result1}) and the dotted lines the numerical data, where $N=500^3$.}
\label{num1}
\end{figure}

In Fig. \ref{num1}, we show the dependence on $u$ of the damping rate (\ref{result1}) at fixed temperatures and the comparison with the numerical calculation without the long-wavelength approximation. As expected, the analytic results well coincide with the numerical data in the vicinity of $u=u_c$ while they deviate as $u$ decreases from $u=u_c$. It is seen that for given $u$ the damping rate monotonically increases as the temperature increases.  This tendency is stronger in a closer vicinity of the critical point $u=u_c$.

%-----------------------------------------------------------------
%-----------------------------------------------------------------

\begin{figure}[htbp]
\begin{center}
\includegraphics[width=80mm]{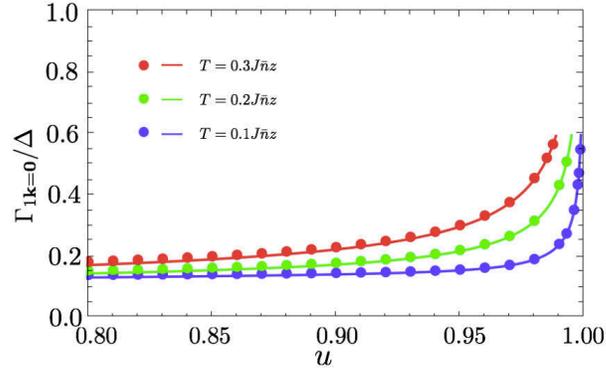}
\end{center}
\vspace{-5mm}
\caption{Dependence on $u$ of the ratio $\Gamma_{1{\bf k=0}}/\Delta$ at fixed temperatures. The solid and dotted lines stand for the analytic and numerical data. The choice of the parameters is the same as that in Fig. \ref{num1}.}
\label{over1}
\end{figure}

The dimensionless quantity $\Gamma_{1{\bf k}={\bf 0}}/\Delta$ characterizes the behavior of the temporal oscillation of the Higgs mode and determines the width of the resonance peak in the spectral function. If $\Gamma_{1{\bf k}={\bf 0}}/\Delta > 1$, then the oscillation abruptly attenuates during one period, i.e., it is overdamped. From Fig.~\ref{over1}, where $\Gamma_{1{\bf k}={\bf 0}}/\Delta$ is plotted against $u$, we see that $\Gamma_{1{\bf k}={\bf 0}}/\Delta$ increases as the critical point is approached but the Higgs mode remains underdamped ($\Gamma_{1{\bf k}={\bf 0}}/\Delta<1$) even at $T=0.3 J\bar{n}z$. Provided the facts that $\bar{n}\gg1$ is assumed and that the temperature can be as low as $T/J=O(1)$ in typical experiments with bosons in optical lattices~\cite{trotzky-10, endres-12}, our result implies that the Higgs mode has a sharp resonance peak in the spectral function at the typical temperatures at least in the absence of a trapping potential. Notice that in Figs.~\ref{num1} and \ref{over1} we did not show the data points in the parameter region, where $T>\Delta \sim T_c$, because our method is valid only in the superfluid phase. Here, $T_c$ denotes the transition temperature from the superfluid to the normal fluid.

%-----------------------------------------------------------------
%-----------------------------------------------------------------

Our calculations do not take into account the logarithmic correction to the damping rate, which stems from the renormalization of an effective coupling constant because $d=3$ is the upper critical dimension~\cite{affleck-92}. Due to the correction, $\Gamma_{1{\bf k}={\bf 0}}/\Delta$ at $T=0$ approaches zero as $\sim 1/\ln|u-u_c|$ in the limit that $u\rightarrow u_c$. The ignorance of the logarithmic correction is not problematic in the practical sense that the parameter region, where the correction is effective, is so narrow that it is very difficult to observe either in experiments or in numerical simulations especially for dynamical quantities like the damping rates.

%%%%%%%%%%%%%%%%%%%%%%%%%%%%%%%
\section{Damping rate of the NG mode}
\label{sec:NG}
%%%%%%%%%%%%%%%%%%%%%%%%%%%%%%%
In this section, we evaluate the damping rate of the NG mode expressed in Eq.~(\ref{damp2}) in specific parameter regions of interest. In cold-atom experiments, the NG mode can be dynamically excited by means of the two-photon Bragg scattering techniques~\cite{huber-07, kozuma-99, stenger-99, ernst-09}, which allows for a wide control of the momentum ${\bf k}$. Hence, we analyze the ${\bf k}$-dependence of the damping rate in addition to the $u$-dependence and the $T$-dependence. First, let us obtain analytical expressions of the damping rate within the long-wavelength approximation. Substituting Eqs.~(\ref{ahg}), (\ref{ang}), (\ref{auv1}), and (\ref{auv2}) into Eq.~(\ref{damp2}) and integrating it with respect to ${\bf k}_1$ and ${\bf k}_2$ except for the variable $|{\bf k}_2|$ lead to
\begin{align}
\Gamma_{2{\bf k}}=-\frac{(1+u)^4(1-u^2)}{8\pi c_{\rm ng}^2c_{\rm h}^2 k^2}\int_{k_l}^{k_u} d|{\bf k}_2| (f_{\rm B}({\cal E}_{2{\bf k}}+{\cal E}_{2{\bf k}_2})-f_{\rm B}({\cal E}_{2{\bf k}_2})),\label{midway}
\end{align}
where $k\equiv|{\bf k}|$. The upper and lower bounds of the integration $k_u$ and $k_l$ are given by
\begin{align}
k_{u}=-k+\sqrt{\frac{\Delta^2}{c_{\rm ng}^2-c_{\rm h}^2}},\hspace{3mm} k_{l}=\frac{1}{c_{\rm ng}^2-c_{\rm h}^2}\left\{-(c_{\rm ng}^2+c_{\rm h}^2)k+\sqrt{4c_{\rm h}^2c_{\rm ng}^2k^2 + \Delta^2(c_{\rm ng}^2-c_{\rm h}^2)}\right\}. \label{lower}
\end{align}
These parameters determine the maximum and minimum momenta of the thermally excited NG modes that are absorbed by the initial NG mode in the Landau damping process.  Carrying out the remaining integration with respect to $|{\bf k}_2|$, we obtain
\begin{equation}
\Gamma_{2 {\bf k}}= \frac{3^{5/2}(1+u)^2(1-u)}{4\beta\sqrt{2}\pi u^2 k^2}{\rm log}\frac{1 - {\rm e}^{-\beta c_{\rm ng}(k_{l}+k)}}{1 - {\rm e}^{-\beta c_{\rm ng}(k_{u}+k)}}\frac{1 - {\rm e}^{-\beta c_{\rm ng}k_{u}}}{1-{\rm e}^{-\beta c_{\rm ng}k_{l}}}. \label{result2}
\end{equation}
We emphasize that this approximate expression (\ref{result2}) is well justified when $k_l \ll 1$. The condition $k_l \ll 1$ can be converted to 
\begin{eqnarray}
z|u-u_c|\ll k, 
\label{NGvalidity}
\end{eqnarray}
meaning that Eq.~(\ref{result2}) is valid only near the critical point.
It is obvious that in the zero-temperature limit ($\beta \rightarrow \infty$) the damping rate (\ref{result2}) vanishes because there is no thermal excitation.

When the conditions that $k_l \ll k_u$ and $c_{\rm ng}k \ll T \ll c_{\rm ng}k_u$ are satified, a further simplification of Eq.~(\ref{result2}) can be made as
\begin{align}
\Gamma_{2 {\bf k}} \approx \frac{3^{5/2}(1+u)^2(1-u)}{4 \sqrt{2}\pi u^2 k} c_{\rm ng}f_{\rm B}( c_{\rm ng}k_l ). \label{result2.2}
\end{align}
Equation (\ref{result2.2}) clearly shows that the damping rate is proportional to the Bose distribution function, $f_{\rm B}( c_{\rm ng}k_l )$, of the thermally excited NG modes with the lowest momentum $k_l$ that is allowed by the conservation law.

\begin{figure}[htbp]
\begin{center}
\includegraphics[width=140mm]{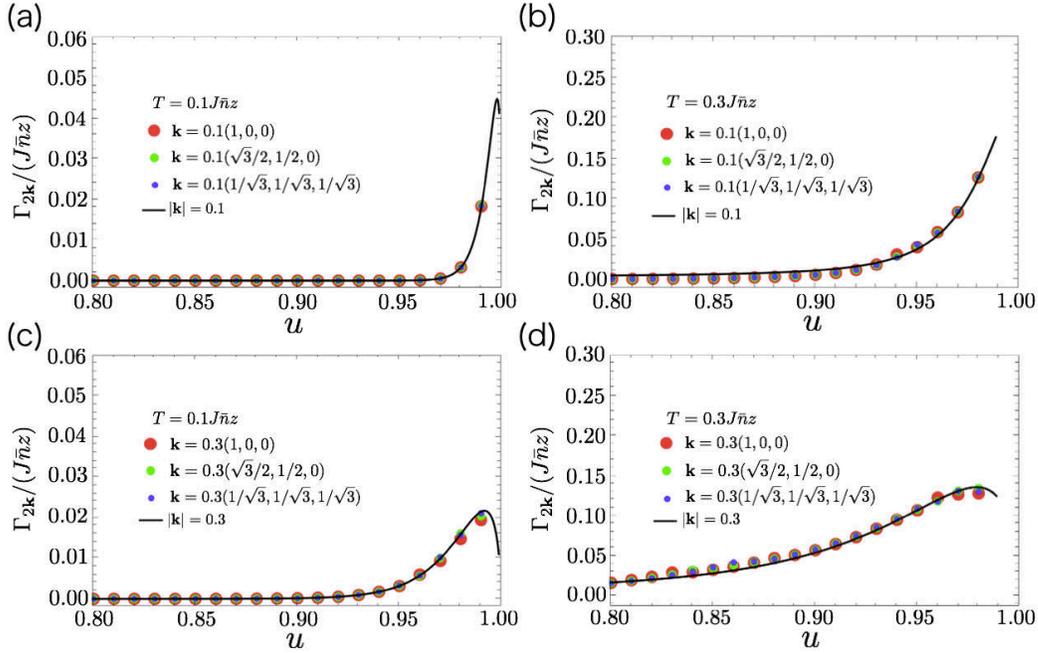}
\end{center}
\vspace{-5mm}
\caption{Dependence on $u$ of the damping rate $\Gamma_{2{\bf k}}/(J\bar{n}z)$ at fixed temperatures and initial momenta. The solid lines represent the analytic results of Eq.~(\ref{result2}) and the dots represent the numerical evaluation of Eq.~(\ref{damp2}) without the long-wavelength approximation. The system size taken in the numerical calculations is $N=500^3$. We set $(T/(J{\bar n}z),k)$ = $(0.1,0.1)$ (a), $(0.3,0.3)$ (b), $(0.1,0.3)$ (c), and $(0.3,0.3)$ (d). Notice that we do not show the region where $T > \Delta \sim T_c$.}
\label{num2}
\end{figure}

In Fig.~\ref{num2}, we show the damping rate of the NG mode $\Gamma_{2{\bf k}}/(J\bar{n}z)$ as a function of $u$ for several values of the temperature $T$ and the initial momentum ${\bf k}$. In addition to the results of the analytical expression (\ref{result2}), we plot the data points obtained from numerical integration of Eq.~(\ref{damp2}) with respect to ${\bf k}_1$ and ${\bf k}_2$ for comparison. In order to clarify effects of the anisotropy of $\gamma_{\bf k}$ in the momentum space, which are ignored in Eq.~(\ref{damp2}), we show the cases of the three different directions, namely ${\bf k}/k=(1,0,0)$, $(\sqrt{3}/2,1/2,0)$, and $(1/\sqrt{3},1/\sqrt{3},1/\sqrt{3})$.

From Eq.~(\ref{result2}) and Fig.~\ref{num2}, we see the following four generic tendencies. First, for given $u$ and ${\bf k}$, the damping rate increases with increasing the temperature. This is very natural in the sense that the number of thermally excited NG modes, which are a main source of the Landau damping, is larger for a higher temperature. Second, the effects of the anisotropy are more noticeable for larger $k$, which is also natural. Third, the analytical results of Eq.~(\ref{result2}) better agree with the numerical data at a closer vicinity of the critical point, as expected from the validity condition (\ref{NGvalidity}). Fourth, it is the most remarkable that the damping rate is significantly large near the critical point and monotonically decays into zero from its peak as $|u-u_c|$ increases.

%-----------------------------------------------------------------
%-----------------------------------------------------------------

\begin{figure}[htbp]
\begin{center}
\includegraphics[width=120mm]{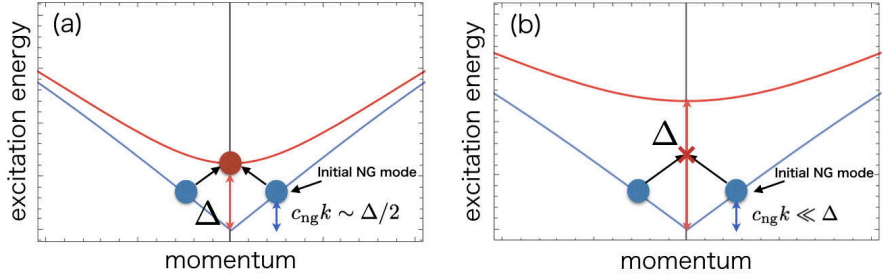}
\end{center}
\vspace{-5mm}
\caption{Schematic illustration of the Landau damping process of the NG mode. The red and blue lines stand for the Higgs and NG branches. The red and blue circles represent the excited Higgs and NG modes. (a) corresponds to a vicinity of the critical point such that $c_{\rm ng}k \simeq \Delta/2$. (b) corresponds to a region far from the critical point such that $c_{\rm ng}k \ll \Delta$.}
\label{pic}
\end{figure}

Let us explain the fourth tendency from a viewpoint of the energy-momentum conservation law. If the Higgs gap satisfies the condition that $c_{\rm ng}k\simeq \Delta/2$ near $u=u_c$ as illustrated in Fig.~\ref{pic}(a), the lowest energy of the thermal NG mode that is absorbed by the initial NG mode is given by $c_{\rm ng}k_l \simeq \Delta/2$ as a result of the energy-momentum conservation law. This means that the Bose distribution function in Eq.~(\ref{result2.2}) can be $f_{\rm B}( c_{\rm ng}k_l )=O(1)$ even at $T<\Delta \sim T_c$. In other words, in this case there are sufficiently many thermal NG modes that the initial NG mode can absorb to attenuate. On the other hand, if $|u-u_c|$ increases for fixed $T$ such that $c_{\rm ng}k\ll \Delta$ [see Fig.~\ref{pic}(b)], the condition that $c_{\rm ng}k_l > \Delta$ is imposed by the energy-momentum conservation law. This implies that $c_{\rm ng}k_l \gg T$, thus leading to the exponential suppression of the damping rate.
%-----------------------------------------------------------------
%-----------------------------------------------------------------

\begin{figure}[htbp]
\begin{center}
\includegraphics[width=140mm]{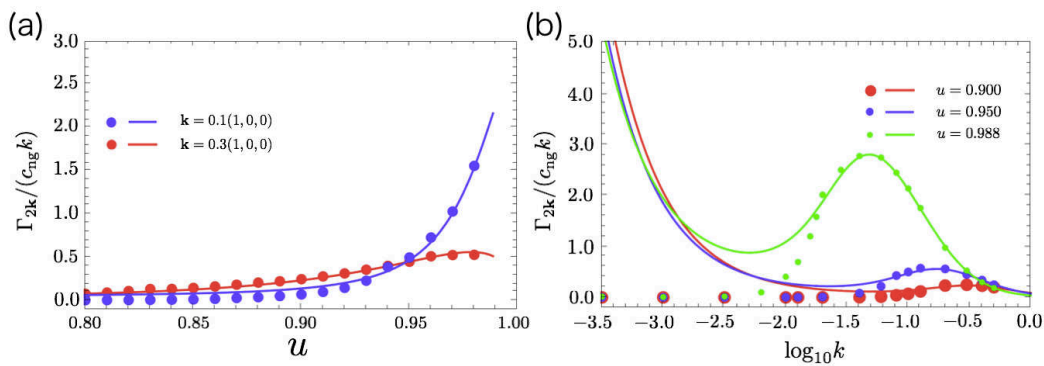}
\end{center}
\vspace{-5mm}
\caption{(a) Dependence on $u$ of the ratio $\Gamma_{2{\bf k}}/(c_{\rm ng}k)$ at $T=0.3J{\bar n}z$ and two values of ${\bf k}$. (b) Dependence on $k$ of the ratio $\Gamma_{2{\bf k}}/(c_{\rm ng}k)$ at $T=0.3J{\bar n}z$, ${\bf k}/k=(1,0,0)$, and three values of $u$. For both cases, we set $N=500^3$ for the numerical calculations.}
\label{num3}
\end{figure}

In Fig.~\ref{num3}, we show the ratio $\Gamma_{2{\bf k}}/(c_{\rm ng}k)$ to characterize the behavior of the temporal oscillation of the NG mode. As seen in Fig.~\ref{num3}(a), when $k=0.1$ and $T=0.3 J\bar{n}z$, $\Gamma_{2{\bf k}}/(c_{\rm ng}k)$ exceeds unity near the critical point, i.e., the NG mode is overdamped. Figure \ref{num3}(b) shows that the damping rate can be even larger by optimizing the wavelength of the initial NG mode $k$. One also sees from Fig.~\ref{num3}(b) that the analytical expression (\ref{result2}) completely fails in the limit of $k\rightarrow 0$, which is consistent with the validity condition (\ref{NGvalidity}).

We finally note the limitation of the 1-loop approximation regarding the predictability of overdamping of a collective mode. Because this approximation is a perturbative approach, $\Gamma_{m{\bf k}}> \mathcal{E}_{m{\bf k}}$, which is the condition of the overdamping, means that the perturbative correction is larger than the nonperturbative value. Therefore, one cannot judge whether or not the overdamping is an artifact of the 1-loop approximation until the higher-order corrections are evaluated.
%-----------------------------------------------------------------
%-----------------------------------------------------------------
%%%%%%%%%%%
\section{Summary}
\label{sec:summary}
%%%%%%%%%%%
In this paper, we studied the damping of the Higgs and the NG modes of Bose gases in a cubic optical lattice at finite temperatures. We calculated the damping rates by using the effective spin-1 model and the field theoretical methods. We derived the analytic expressions of the damping rates within the long-wavelength approximation and confirmed their validity in the vicinity of the critical point through the comparison with numerical calculations. We showed that while the Higgs mode attenuates more significantly at higher temperatures, it is not overdamped at temperatures that can be achieved in typical experiments. This result indicates the feasibility of detecting the Higgs mode at 3D and the finite temperatures as a resonance peak in a spectral function at least when there is no trapping potential. As for the NG mode, we found parameter regions, where the Landau damping process leads to the overdamping of the NG mode, and discussed the origin of the strong damping especially near the critical point.

In the future studies, it will be important for addressing the detectability of the Higgs mode in 3D more quantitatively to include effects of the trapping potential and the breaking of the particle-hole symmetry. While a prescription to treat these effects within the tree level has been presented in Ref.~\cite{huber-07}, one needs to extend it to the 1-loop level to take into account the effects of quantum and thermal fluctuations as well.

\section*{Acknowledgment}
The authors thank S. Tsuchiya for useful discussions. The authors also thank the Yukawa Institute for Theoretical Physics (YITP) at Kyoto University, where this work was initiated during the YITP workshop (YITP-W-14-02) on "Higgs Modes in Condensed Matter and Quantum Gases". I.~D. acknowledges Grants-in-Aid for Scientific Research from JSPS: Grants No.~25800228 and No.~25220711.

\end{document}